\documentclass[aps,prl,twocolumn]{revtex4}
\usepackage{amsmath}
\usepackage{amssymb}

\newcommand\ro{{\Hat\rho}}
\newcommand\sov{\mbox{\boldmath{$\Hat\sigma$}}}
\newcommand\av{\mathbf{a}}
\newcommand\bv{\mathbf{b}}
\newcommand\br{\mathbf{r}}
\newcommand\Io{{\Hat 1}}
\newcommand\dd{\mathrm{d}}
\newcommand\tr{\mathrm{tr}}

\newcommand\M{\mathrm{M}}

\begin{document}
\title{Shortnote on local hidden Grassmann variables vs. quantum correlations}
\author{Lajos Di\'osi}
\email{diosi@rmki.kfki.hu}
\homepage{www.rmki.kfki.hu/~diosi} 
\affiliation{
Research Institute for Particle and Nuclear Physics\\
H-1525 Budapest 114, POB 49, Hungary}
\date{\today}

\begin{abstract}
Grassmannian local hidden variables are shown to generate all possible quantum correlations 
in a bipartite quantum system. Grassmann representation of fermions, common in field theory, 
opens a related perspective. Although Grassmann hidden variables can not challange Bell's 
locality theorem, they can become an interesting mathematical tool to investigate entanglement. 
\end{abstract}

\maketitle
Given two uncorrelated quantum systems $A$ and $B$ with density matrices $\ro_A$ and
$\ro_B$, respectively, the state of the bipartite composite system is the tensor product 
$\ro_A\ro_B$. Let the states of both $A$ and $B$ depend on a certain variable $\lambda$ so 
that the composite state were $\ro_B(\lambda)\ro_B(\lambda)$ had we known the value of 
$\lambda$. However, we suppose that $\lambda$ is {\it hidden} variable in the sense that
we only know the statistics of it. Therefore, the emerging composite state is the 
statistical mean value of $\ro_B(\lambda)\ro_B(\lambda)$:
\begin{equation}\label{sep}
\ro_{AB}=\M[\ro_A(\lambda)\ro_A(\lambda)]
\end{equation}
defined through the normalized probability $p$ of the hidden variable:
\begin{equation}\label{stat}
\M[\dots]=\int\dots p(\lambda)\dd\lambda~.
\end{equation}
After Werner \cite{Wer89}, we consider eq.~(\ref{sep}) the {\it separability} condition for the state $\ro_{AB}$.  
If $A$ and $B$ were classical systems their composite states would always be separable, i.e., each composite
classical density is a weighted mixture of uncorrelated densities. We say classical correlations emerge from 
ignorance regarding some hidden variables. This is not so in quantum theory. The separable quantum states which are 
mixtures of uncorrelated states will be called classically correlated quantum states.
The non-separable quantum states $\ro_{AB}$, for which the expansions (\ref{sep}) do not exist, are called
quantum correlated or, equivalently, entangled. The existence of non-classical
correlations is a principal difference of quantum theory from the classical one.

The lack of separability (\ref{sep}) shows up for two Pauli spins $\sov_A$ and $\sov_B$ already.    
The most general forms of the two spin states, respectively, read: 
\begin{equation}\label{rhoab}
\ro_A(\av)=\frac{1}{2}(\Io_A+\av\sov_A)~,~~~
\ro_A(\bv)=\frac{1}{2}(\Io_A+\bv\sov_B)~,
\end{equation}
where $\av=(a_1,a_2,a_3)$ and $\bv=(b_1,b_2,b_3)$ are real spatial polarization vectors satisfying $\av^2,\bv^2\leq1$.
Without restricting generality, the hidden variable is the pair of the polarization vectors, $\lambda=(\av,\bv)$, with 
the probability distribution $p(\av,\bv)$.
The composite state $\ro_{AB}$ is separable if the probability distribution $p(\av,\bv)$ exists such that
\begin{equation}\label{sepab}
\ro_{AB}=\M[\ro_A(\av)\ro_B(\bv)]~.
\end{equation}
Let us apply this condition to the rotational invariant special case where $\M[\av]=\M[\bv]=0$ and, in particular,
\begin{equation}\label{eta}
\M[a_ib_j]=\eta\delta_{ij}
\end{equation}
for $i,j=1,2,3$. Most importantly, the correlation $\eta$ is constrained by $\vert\eta\vert\leq1/3$
since the values of $\av$ and $\bv$ were constrained by $\av^2,\bv^2\leq1$.
If we substitute (\ref{rhoab}) and (\ref{eta}) into the separability condition (\ref{sepab}), we get the following form:
\begin{equation}\label{rhoABeta}
\ro_{AB}=\frac{1}{4}(\Io_A\Io_B+\eta\sov_A\sov_B)~,
\end{equation}
which is thus separable if $\vert\eta\vert\leq1/3$ and non-separable otherwise \cite{Wer89}.
The matrix $\ro_{AB}$ is non-negative for $\eta\in[-1,1/3]$ hence the states are indeed 
non-separable (quantum correlated, entangled) for $\eta\in[-1,-1/3)$. 
The constraints $\av^2,\bv^2\leq1$ have forbidden the existence of
a hidden variable probability distribution $p(\av,\bv)$ that could provide the (anti-)correlation
stronger than $\eta=-1/3$. Quantum mechanics can achieve $\eta=-1$ as well. 
To generate stronger than $\eta=-1/3$ anti-correlations via hidden variables, 
we could adopt the blunt compromise as to allow $p(\av,\bv)\not\geq0$ 
which means that we would give up the statistical interpretation
of the hidden variables. 

We take an even more radical step instead: in the
separability condition (\ref{sep}) we assume formally 
that the hidden classical variable $\lambda$ is Grassmann variable. As for eq.~(\ref{stat}), the theory
of Grassmann variables contains the notion of integral \cite{Ber66} and of the corresponding measure $p(\lambda)$.

In particualar, we are going to discuss the case of the two correlated Pauli spins. 
Suppose $\av,\bv$ are Grassmann variables:
\begin{equation}
a_i a_j+a_j a_i=b_i b_j+b_j b_i=a_i b_j+b_j a_i=0~,
\end{equation}
for all $i,j=1,2,3$. 
Let us choose the following normalized rotation invariant Gaussian distribution \cite{Ber66} of the Grasmannian 
hidden variables:
\begin{equation}\label{pGr}
p(\av,\bv)\dd\bv\dd\av=\eta^3\mathrm{exp}\left(\eta^{-1}\av\bv\right)\dd\bv\dd\av~,
\end{equation}
satisfying $\M[\av]=\M[\bv]=0$ and the correlation equation (\ref{eta}). Since $\eta$ is not constrained at all,
we conclude that the Grassmann hidden variables $\av,\bv$ can generate all possible correlations that two
Pauli spins may have in quantum mechanics. 

It is plausible to conjecture that quantum correlations can universally be reproduced by Grassmann hidden variables. 
Namely, a multipartite composite state can be expressed in the form
\begin{equation}
\ro_{AB\dots K\dots}=\M[\ro_A(\lambda)\ro_B(\lambda)\dots\ro_K(\lambda)\dots]
\end{equation}
where $\lambda$ is the hidden variable. If the state is separable then $\lambda$ can be chosen real valued; 
if the state is entangled then $\lambda$ must be a combination of real and Grassmann numbers. 
Perhaps the real numbers generate the classical while the Grassmann ones generate the quantum correlations, respectively, 
although the existence of such separation is an open issue itself. The combination of real number and Grassmann algebras 
might give some new insight into the generic entanglement structure.

I got the hint of Grassmann hidden variables from quantum field theory where the  
fermionic quantum fields $\Hat\psi(t,\br)$ can equivalently be represented by the corresponding 
local Grassmann fields $\psi(t,\br)$ that satisfy a certain normalized distribution \cite{Ber66}. 
This means, at least for equilibrium states, that all fermion-mediated quantum correlations (entanglements) 
emerge from the local Grassmann variables $\psi(t,\br)$ which play the role of hidden variables.

Another motivation comes from the recent work \cite{Chr07} by Christian, who suggested Clifford algebra valued hidden 
variables to violate Bell inequalities. In both his and my proposal the non-commutative hidden variables are able to 
generate entanglement of the composite state. Grassmann might have some theoretical advantage 
over Clifford numbers because of the mentioned universal Grassmann-fermion correspondence. Therefore I see a
certain mathematical perspective to treat entanglement in the language of Grassmann hidden variables.
The present work has no intention to challange the Bell theorem \cite{Bel64} since Grassmann (or Clifford) numbers 
can not parametrize individual measurement results \cite{foot}, only real valued hidden variables can, cf. \cite{Gra07}. 
Yet, the proposed hidden Grassmann variables $\lambda$ are local in the spirit close to Bell's: 
they are by construction independent of the occasional local experimental settings at sides $A$ or $B$.   

This work was supported by the Hungarian OTKA Grant No. 49384.

\vfill

\begin{thebibliography}{99}
\bibitem{foot} E.g., to the measurement of $\Hat\sigma_{A3}$ the eq.~(\ref{rhoab}) 
would assign the value $\tr(\Hat\sigma_{A3}\ro_A)$ which is the Grassmannian $a_3$.
\end{thebibliography}
\end{document}